\documentclass{article}

\usepackage{amsmath}  

\newtheorem{theorem}{Theorem}[section]

\newtheorem{definition}[theorem]{Definition}
\newtheorem{problem}[theorem]{Problem}

\newcommand\php{{\mbox{PHP}}}
\newcommand\np{{\cal N}{\cal P}}
\newcommand\bits{\{0,1\}}

\begin{document}

\title{The Cook-Reckhow definition}

\author{Jan Kraj\'{\i}\v{c}ek}

\date{Faculty of Mathematics and Physics\\
Charles University\thanks{Sokolovsk\' a 83, Prague, 186 75,
The Czech Republic, {\tt krajicek@karlin.mff.cuni.cz}}}

\maketitle

The Cook-Reckhow paper \cite{CooRec} introduced the notions of {\em propositional proof systems} and
{\em polynomial simulations} among them, described several classes of logical propositional calculi
and compared them with regards to their efficiency, and introduced the {\em pigeonhole principle tautology}
$\php_n$ that is the prime example of a tautology hard to prove in weaker systems ever since, rivaled only by the
tautology proposed earlier by Tseitin \cite{Tse}. 
It also showed that the central
question whether there exist a propositional proof system allowing polynomial size 
proofs of all tautologies is equivalent to a central
question of complexity theory whether the class $\np$ is closed under complementation.

Classical proof theory of first-order logic developed in the first half of the twentieth century
assigned to proofs several combinatorial characteristics and some of them
can be perceived as measures of complexity; for example, the height of a proof tree.
Primary emphasis was on constructions producing various normal forms of proofs and
the combinatorial characteristic helped to measure the progress of normalization constructions.
The question about the minimum length of a proof of a statement (measured by either the number of steps or by
the size, i.e. the number of symbols) was also studied, primarily in the context of speed-up results; references
\cite{God36,Mos52,EhrMyc,Par71,Par73} can serve as illustrations of this research. The results rest
on constructions underlying the undecidability of the Halting problem or G\"{o}del's Incompleteness theorem,
and do not give any insight\footnote{In fact, Parikh \cite{Par71} introduced 
theory PB (called now $I\Delta_0$, cf.\cite{kniha,prf})
which later in the 1980s turned out to be important for the development of proof complexity.} 
into analogous problems in propositional logic.

It was the Cook-Reckhow 1979 paper \cite{CooRec} which
defined the area of research we now call proof complexity. There were earlier papers which contributed to
the subject as we understand it today, the most significant being Tseitin's \cite{Tse}.
But none of them introduced general notions\footnote{Tseitin's paper \cite{Tse} offers no motivation for the
research reported there but one of the motivations were questions we now formulate as the $\cal P$ vs. $\np$
problem (another motivation was computer processing of natural languages) and the special role the 
{\em Entscheidungsproblem for propositional calculus} plays in them (personal communication).}
that would allow to make an explicit and universal link between 
lengths-of-proofs problems and computational complexity theory.

In this note we shall highlight three particular definitions from the paper: of proof systems, p-simulations and 
the formula $\php_n$, and discuss their role in defining the field. We will also mention some related 
developments and open problems.
In particular, we shall show that the general definition of proof systems that has seemingly 
little to do with how ordinary logical calculi are defined is actually equivalent to the calculi definition
with a more general treatment of logical axioms than is usual (Section \ref{def}), we shall present 
the optimality problem stemming from the notion of simulations (Section \ref{sim}), 
and we shall discuss the role of the $\php_n$ formula in proof complexity lower bounds, its limitations and
modern variants aimed at strong proof systems (Section \ref{hard}). Paper \cite{CooRec} also discusses
few measures of complexity of proofs other than the proof size and describes some relations among them; 
we shall not discuss this and instead we refer the reader to available literature.

The Cook-Reckhow paper \cite{CooRec} had a precursor \cite{CooRec74}, an extended abstract that summarized
some research presented later in \cite{CooRec}, as well as from Reckhow's PhD Thesis \cite{Rec}. 
This earlier paper differs from \cite{CooRec} in several
aspects: it uses simulations (see Sec.\ref{sim}) as opposed to the latter finer notion of p-simulations,
and it contains neither Extended Frege system nor the
$\php$ tautology. It presents a rather succinct version of the construction underlying
Reckhow's theorem that was replaced in \cite{CooRec} by a similar statement for Extended Frege systems
with much easier (and more illuminating) proof (cf. Sec.\ref{sim}). It also treats in detail the sequent calculus 
that is only glanced over in \cite{CooRec}, and it derives some super-polynomial 
lower bounds, using Tseitin's results in \cite{Tse},
for weak proof systems such as tree-like resolution or semantic trees.

The aim of this note is to give an idea to the non-expert reader about the main ideas stemming from \cite{CooRec}
and about the fundamental problems of proof complexity. 
Further information about proof complexity, its basic as well as advanced parts,
and about topics in mathematical logic and complexity theory it relates to can be found in \cite{kniha,prf}.

\section{Definition of proof systems} \label{def}

The main example of a logical propositional calculus to keep in mind is a {\em Frege system}.
It is any calculus operating with propositional formulas over a complete basis of logical
connectives (i.e. all Boolean functions can be defined in the language), having a finite
number of sound axiom schemes and inference rules that are {\em implicationally complete}.
The latter term means that if a formula $A$ is a logical consequence of formulas $B_1, \dots, B_m$
then it can be derived from them in the calculus. An example of a complete language is the DeMorgan language
with constants $0,1$ and connectives $\neg, \vee$ and $\wedge$. We shall denote by TAUT the set of tautologies
in this language and we shall tacitly assume that $\mbox{TAUT} \subseteq \bits^*$, with formulas being
encoded by binary strings in some natural way.

There is a number of such systems described in logic text-books and they are often called {\em Hilbert-style},
referring to Hilbert's work in proof theory \cite{HilAck,HilBer34,HilBer39}.
The form of calculi is based on Frege's \cite{Fre79}, hence the name Cook and Reckhow \cite{CooRec} chose
for this class of propositional calculi.

The calculi are {\em sound} (every provable formula is a tautology) and {\em complete} (every tautology is
provable). In addition, the key property singled out by \cite{CooRec} is that to recognize whether a string
of symbols is a valid proof in the calculus or not is computationally feasible: it can be done by a p-time algorithm.
This leads to the following fundamental definition.

\begin{definition} [Cook-Reckhow \cite{CooRec}] \label{pps}
{\ }

A {\bf propositional proof system} is any p-time computable function 
$$
f\ : \bits^* \rightarrow \bits^*
$$
such that
$$
\mbox{TAUT}\ =\ Rng(f)\ .
$$
Any $w \in \bits^*$ such that $f(w) = A$ is called an {\bf $f$-proof of $A$}.
\end{definition}
Cook and Reckhow \cite[Def.1.3]{CooRec} actually define more generally a proof system for any $L \subseteq \bits^*$
by the condition $L = Rng(f)$, and consider proof systems for the set of tautologies in any fixed language.

A Frege system $F$ can be represented by a function $f$ which takes a string $w$ and
maps it to the last formula of $w$, if $w$ is a sequence of formulas that forms a valid $F$-proof, or to constant $1$ if
$w$ is not an $F$-proof. The soundness of $F$ implies that $Rng(f) \subseteq \mbox{TAUT}$ and
its completeness implies the opposite inclusion $\mbox{TAUT} \subseteq Rng(f)$.

It is easy to see that a number of other classes of propositional calculi considered in mathematical logic literature
fit the definition in the same sense as Frege systems do.
These calculi include resolution, sequent calculus or natural deduction. 
Less usual examples of propositional proof systems can be constructed as follows. Take a
consistent first-order theory axiomatized by a finite number of axioms and axiom schemes that is
sound and contains some simple base theory (in order to guarantee both the correctness and the completeness)
and interpret it as a proof system: a proof of 
formula $A$ is a proof in the theory of the formalized statement $A \in \mbox{TAUT}$.
Yet another examples are logic calculi that are set-up to prove the unsatisfiability of formulas: these can 
be interpreted as proof systems by accepting a refutation of $\neg A$ as a proof of $A$.

In addition, the general form of the definition allows us to interpret various calculations in
algebra as propositional proofs. Here it is more natural to speak about refutation systems. If we
have a CNF formula that is a conjunction of clauses $C_i$, we can represent each $C_i$
by a constraint of an algebraic form and use a suitable algebraic calculus to derive the
unsolvability of the formula. For example, 
a clause
$$
p \vee \neg q \vee r
$$
together with the requirement that we look for $0-1$ solution
can be represented by polynomial equations
$$
(1-p)q(1-r)=0\ ,\ p^2-p = 0\ ,\ q^2 -q = 0\ ,\ r^2 - r = 0 
$$
the first equation states that the clause contains a true literal while
the last three equations force $0-1$ solutions over any integral domain. In this case we can use a calculus
deriving elements of the ideal generated by the equations representing similarly all clauses of the formula,
trying to
derive $1$ as a member of the ideal and thus demonstrating the unsolvability of the equations and hence the
unsatisfiability of the formula.

Another approach is to represent the clause as integer linear inequalities
$$
p + (1-q) + r \geq 1\ ,\ 1 \geq p, q, r \geq 0 
$$
and use some integer linear programing algorithm to derive the unsolvability of the system of inequalities
representing the whole CNF formula.
It is a great advantage of Definition \ref{pps} that it puts all these quite different formal system
under one umbrella.

Proof systems can be also defined equivalently in a relational form. A {\em relational propositional proof system}
is a binary relation $P(x,y)$ that we interpret as the provability relation {\em $y$ is a proof of $x$}. It is
required that it is p-time decidable and that for any formula $A$ it holds:
$$
A \in \mbox{TAUT}\ \mbox{ iff }\ 
\exists w P(A,w)\ .
$$
This is closer in form to logical calculi (and can be represented by the function version as Frege systems
were before) but it is equally general: a functional proof system $f$ is represented by the relation $f(y)=x$.

A proof system $f$ is {\bf p-bounded} iff there exists $c \geq 1$ such that for all $A$, $|A| > 1$,
$$
A \in \mbox{TAUT}\ \Rightarrow \ 
\exists w (|w| \le |A|^c)\ f(w) = A\ .
$$
In the relational form this would read
$$
A \in \mbox{TAUT}\ \Rightarrow \ 
\exists w (|w| \le |A|^c)\ P(A,w)
$$
and combining this with the soundness we get
$$
x \in \mbox{TAUT}\ \Leftrightarrow\ \exists y (|y| \le |x|^c) P(x,y)\ .
$$
The right-hand side expression has the well-known general form in which any $\np$ set can be defined.
Hence we get as a simple but important corollary to the definition the following statement (the second
equivalence uses Cook's theorem: the $\np$-completeness of SAT, cf.Cook \cite{Coo71}).

\begin{theorem} [Cook - Reckhow \cite{CooRec}] \label{21.8.19a}
{\ }

A p-bounded proof system exists iff $\mbox{TAUT} \in \np$ iff $\np = co\np$.
\end{theorem}
This theorem determines  
\begin{problem} [Main problem of proof complexity]
{\ }

Is there a p-bounded proof system for TAUT?
\end{problem}
By Theorem \ref{21.8.19a} showing that no p-bounded proof system exists would imply, in particular, 
that ${\cal P} \neq \np$ because $\cal P$ is closed under complementation. 
On the other hand, defining a p-bounded proof system $f$
would allow to witness various $co\np$-properties by short witnesses ($f$-proofs); \cite{CooRec74} mentions
the property that two graphs are not isomorphic.

One may consider variants of the definition of proof systems when the provability relation is not necessarily decidable
by a p-time algorithm but only by more general algorithm; for example, using some randomness. My view is that
this changes the basic problems of proof complexity substantially. While it may link propositional proof systems
with various other proof systems considered in different parts of complexity theory, it is not clear
that it will shed light on proof complexity proper. 
This may change if some of these other parts of complexity theory
advance significantly on their own fundamental open problems.

The Cook-Reckhow definition is handy for establishing Theorem \ref{21.8.19a} and
the connection to complexity theory but the reader may wonder if it does not deviate from logical form 
of calculi too much. In fact, it can be shown that every proof system 
can be p-simulated (in the sense of the next section) by a Frege system whose set of axioms is not given just by
a finite number of axiom schemes but is possibly infinite but easy to recognize (in p-time, in particular) sparse subset of TAUT.
Doing this precisely is rather technical and we refer the reader to \cite{KraPud89a,kniha,prf}.

\section{Simulations among proof systems} \label{sim}

When studying the problem whether some proof system is p-bounded it is useful to be able to compare two proof
systems with respect to their efficiency. The following two notions\footnote{P-simulations are
also defined in Cook \cite{Coo75}.} are aimed at that.

\begin{definition} [Cook-Reckhow \cite{CooRec}]
{\ }

Let $f, g$ be two proof systems. A {\bf simulation} of $g$ by $f$ is any function
$$
h\ :\ \bits^* \rightarrow \bits^*
$$
such that for all $w \in \bits^*$, $|h(w)| \le |w|^c$, for some independent constant $c \geq 1$ and all $|w| > 1$, 
and such that
$$
f(h(w))\ =\ g(w)\ .
$$
Simulation $h$ is {\bf p-simulation} if it is p-time computable.

Proof system $f$ {\bf (p-)simulates} $g$ ($f \geq g$ and $f \geq_p g$ in symbols, respectively) iff
there is a (p-)simulation of $g$ by $f$.
\end{definition}
In other words, the statement that $f \geq g$ says that if we replace $f$ by $g$ we can speed-up proofs at most polynomially,
while the statement that
$f \geq_p g$ says that we can even efficiently translate $g$-proofs into $f$-proofs. Both these relations are quasi-orderings
(we get partial orderings after factoring by the equivalence relations of mutual simulations).

There are other options how to define a quasi-ordering of proof systems. In particular, if we did not insist
in Definition \ref{pps} that all proof systems prove tautologies in the same language (we have defined TAUT using
the DeMorgan language only)
but allowed tautologies in different languages then a (p-)simulation should allow to translate also formulas and
not just proofs. 
By insisting that the target set is TAUT we forced that such a translation of formulas is incorporated into
the definition of particular proof systems that may operate with formulas in other languages or even with polynomials
or other objects.
In fact, considering instead of propositional proof systems proof systems for
any $co\np$-complete set we ought to
allow p-reductions between such sets and TAUT. 

However, for positive results (as is Theorem \ref{19.8.19a} bellow)
p-simulations allow to formulate the strongest possible statements while strongest negative results (obtained by proving
a super-polynomial lower bound for $f$-proofs of formulas for which there are polynomial size $g$-proofs) talk about 
super-polynomial speed-ups and hence about the non-existence of simulations. Thus the two types of simulations
serve their purpose very well.

Cook and Reckhow \cite{CooRec} compared various logical proof systems in terms of p-simulations; the following statement
summarizes their most memorable results in this respect.

\begin{theorem} [Cook-Reckhow\cite{CooRec}] \label{19.8.19a}
{\ }

\begin{enumerate}

\item All Extended Frege systems in all languages p-simulate each other.

\item Frege systems and propositional parts of natural deduction and of sequent calculus 
mutually p-simulate each other.

\item Extended Frege system EF and Tseitin's Extended resolution ER are p-equivalent and they are p-simulated
by any Frege system with the substitution rule.

\end{enumerate}
\end{theorem}
Extended Frege systems EF were defined in \cite{CooRec} in a direct analogy with Extended resolution ER of
Tseitin \cite{Tse}. Any such system starts with a Frege system and allows, in addition, 
to abbreviate formulas by new atoms and use these in proofs. In particular,
during an EF-proof we can take a new atom $q$ (an {\em extension atom}) not used so 
far and not occurring in the target formula $A$
to be proved, any formula $D$ not containing $q$, and introduce the equivalence $q \equiv D$ (represented in the
language of the system) as a new {\em extension axiom}. 
Note that EF is not a Frege system as the introduction of extension axioms does not fit the schematic way Frege axioms
are supposed to be defined.
The first statement in the theorem
is a weaker version of Reckhow's theorem \cite{Rec} which is stated for Frege systems. The version for Extended Frege
system is much easier to prove (see \cite{kniha,prf} for published proofs of the stronger version). 

For the definition of natural deduction see \cite{Pra},
for sequent calculus see any of \cite{Gen,kniha,prf} (the sequent calculus part of the statement is just mentioned 
in \cite{CooRec} while natural deduction is treated in detail).
The substitution rule allows to infer from a
formula $B(p_1, \dots, p_m)$ its arbitrary substitution instance $B(C_1, \dots, C_m)$ in one inference. A Substitution Frege
system SF is a Frege system augmented by this rule. It was proved later in \cite{Dow79} (indirectly) and in \cite{KraPud89a}
(an explicit p-simulation) that EF actually p-simulates SF as well.

\bigskip

An illuminating description of EF is that it is essentially a Frege system that operates with circuits rather 
than with formulas; this has been made precise in \cite{Jer04}. Perhaps even more useful is the  
statement that the minimum size $s$ of an EF-proof of formula $A$ is proportional 
to the minimum number of steps in a Frege proof of $A$ and $|A|$, or to the minimum 
number of different formulas that need to occur as subformulas in any Frege proof of $A$ and $|A|$, cf. 
\cite{CooRec} or \cite{kniha,prf}. Hence moving from F to EF means that we are replacing the size as the
measure of complexity of Frege proofs by
the number of steps. This is interesting because from the point of view of mathematical logic
the number of steps is a very natural complexity measure. 

Extended Frege system is also important because of its relation to a particular theory PV introduced by
Cook \cite{Coo75} at the same time (he used ER in his paper). 
This is discussed in S.~Buss's article in this volume.
Theory PV (stands for Polynomially Verifiable) allows to formalize a number 
of standard computational complexity constructions and arguments.
Understanding the power of proof system EF and, in particular, showing that it is not p-bounded, is considered
in the field as the pivotal step towards solving the Main problem and 
proving that $\np \neq co\np$. In particular, it is also known
that any super-polynomial lower bound for EF implies that $\np \neq co\np$ is consistent with PV (cf. \cite[Sec.12.4]{prf}).

We shall mention one problem formulated only later in \cite{KraPud89a} which is, however, natural
and is implicit in the definition of simulations.

\begin{problem} [Optimality problem]
{\ }

Is there a proof system that (p-)simulates all other proof systems?
\end{problem}
Such a maximal proof system is called {\bf (p-)optimal} after \cite{KraPud89a}. 
We have (names for) three types of proof systems whose existence is considered by most researchers unlikely:
p-bounded, p-optimal and optimal. Every p-bounded or p-optimal proof system is also optimal
and this rules out three out of eight possibilities for the existence/non-existence of objects of these three types.
At present we cannot rule out any of the remaining five scenarios: 
\begin{itemize}

\item {\em A p-bounded, p-optimal proof system $P_1$ exists.}

Having such an ideal proof system we do not need to consider any other: even searching for proofs
in any other proof system can be reduced to searching for $P_1$-proofs. (We ignore here that p-reductions
themselves increase polynomially the time complexity of a proof search algorithm and 
may transform a combinatorially transparent one into a complex one, cf. the last paragraph of this section.)

\item {\em A p-bounded proof system $P_2$ exists but no p-optimal does.}

While p-size $P_2$-proofs would exist for each tautology, finding them may be
difficult and it may help to consider different proof systems for different (classes of)
tautologies.

\item {\em A p-optimal proof system $P_3$ exists but no p-bounded does.}

Here we can restrict our attention to $P_3$: it is also optimal and search for proofs in any proof system
can be replaced by a search for $P_3$-proofs.

\item {\em An optimal proof system $P_4$ exists but no p-bounded or p-optimal does.}

Proving lengths-of-proofs lower bounds (or upper bounds, for that matter) can be restricted to $P_4$
but proof search may benefit from considering different proof systems for different classes of tautologies.

\item {\em None of these ideal objects exist.}
 
This appears to be the most likely scenario.

\end{itemize} 
At present we cannot rule out that a Frege system is one of $P_1, \dots, P_4$.
The Optimality problem is related to a surprising number of varied topics in proof theory (quantitative G\"{o}del's
theorem), finite model theory, structural complexity, and some other (cf. \cite[Chpt.21]{prf}).

An interesting question left-out by \cite{CooRec} as well as in later literature is how to compare proof search 
algorithms. A tentative definition was proposed in \cite[Sec.21.5]{prf}.

\section{Hard tautologies and the $\php_n$ formula} \label{hard}

In order to prove lengths-of-proofs lower bounds for a proof system we start with a suitable candidate tautology
that we conjecture to be hard to prove (i.e. requiring long proofs) therein.
A particular tautology for this purpose based on the pigeon-hole principle was proposed in \cite{CooRec}. 
The formula, to be denoted $\php_n$, 
is built from atoms $p_{i j}$ with $i \in [n]:=\{1, \dots, n\}$ and $j \in [n-1]$, for
$n \geq 2$. Thinking of $p_{i j}$ as representing the atomic statement that {\em $i$ maps to $j$}, we
can express that the map is defined at $i$ by the clause
\begin{equation} \label{20.8.19a}
\bigvee_j p_{i j}
\end{equation}
the fact that $j$ can be the value of at most on $i$
by
\begin{equation} \label{20.8.19b}
\bigvee_{i_1 \neq i_2} \neg p_{i_1 j} \vee \neg p_{i_2 j}
\end{equation}
and the fact that $i$ maps to at most one value by
\begin{equation} \label{20.8.19c}
\bigvee_{j_1 \neq j_2} \neg p_{i j_1} \vee \neg p_{i j_2}\ .
\end{equation}
Taking the conjunction of these clauses for all choices of $i$ and $j$ states that 
$$
\{(i,j) \in [n] \times [n-1]\ |\ p_{i j} =1\}
$$
is the graph of an injective map from $[n]$ into $[n-1]$. No such map exists and hence the negation of the
conjunction is a tautology. This leads to the following definition.

\begin{definition} [Cook-Reckhow \cite{CooRec}]
{\ }

For any $n \geq 2$, $\php_n$ is the disjunction of negations of clauses in (\ref{20.8.19a}) for all $i \in [n]$ 
and in (\ref{20.8.19b}) for all $j \in [n-1]$ and in (\ref{20.8.19c}) for all $i \in [n]$.
\end{definition}
In fact, to reach a contradiction we do not need the assumption that it is the graph of a function,
a multi-function suffices (if $i$ occupies more values $j$ it is harder to be injective). In other words, we do not 
need to include the clauses from (\ref{20.8.19c}) and \cite{CooRec} did not included them. Nowadays the definition
of $\php_n$ as formulated above is more customary and proving lower bounds for it yields stronger results than for
the more economical version (the principle assumes more and hence it is logically weaker). 

Cook and Reckhow \cite{CooRec} showed that it is possible to prove $\php_n$ 
in Extended Frege systems by a proof of size polynomial in $n$ (note that the size of $\php_n$
is also polynomial in $n$).
In fact, they introduced EF in order to formalize smoothly the inductive argument:
from an assignment violating $\php_n$ we can define (using the extension rule)
an assignment violating $\php_{n-1}$.
Hence $\php_n$ has also a proof in Frege systems with
a polynomial number of steps (but having large size).  Buss \cite{Bus-php} improved the result (by a different construction
formalizing counting) and proved that Frege systems actually also admit polynomial size proofs of $\php_n$. 

On the other hand, in a breakthrough result, 
Haken \cite{Hak85} proved a first lower bound for resolution using $\php_n$ 
and the same formula was proved to be hard for constant depth subsystems of any Frege system
in the DeMorgan language by Ajtai \cite{Ajt88} (Haken's lower bound was exponential while Ajtai's super-polynomial - its
rate was later improved to exponential too by \cite{KPW,PBI}). 
The same formula (represented by polynomial equations similarly as
in Section \ref{pps}) served to Razborov \cite{Raz98} for his lower bound for polynomial calculus,
an algebraic proof system manipulating polynomials.

There is an important variant of the PHP formula considered first by Paris, Wilkie and Woods \cite{PWW}
in the context of bounded arithmetic: allow $i$ to range over a much bigger set than $[n]$; for example, over
$[2n]$ or even $[n^2]$. Similarly as the PHP principle is related to counting 
these {\em weak} PHP principles relate to approximate counting and \cite{PWW} showed that they can be
sometimes used in place of PHP proper and that, crucially, they are easier to prove. Their proof (formulated using
bounded arithmetic) gives quasi-polynomial size proofs in constant depth Frege systems of formulas
formalizing these weaker principles for $n \geq 2$.

Even if the formula $\php_n$ itself cannot be used as a hard example for proof systems like F or EF, formulas
formalizing a form of a weak PHP in a different way possibly can. It has been an insight of Wilkie (result reported in \cite[Sec.7.3]{kniha})
that the {\em dual weak PHP} for p-time functions is important in bounded arithmetic (this has been much 
extended by Je\v r\'{a}bek \cite{Jer04,Jer-apc1,Jer-apc2}). The principle says that
no p-time function $g$ when restricted to any $\bits^n$ can
can be onto $\bits^{2n}$. Now take an arbitrary $b \in \bits^{2n} \setminus Rng(g_n)$, where
$g_n$ is the restriction of $g$ to $\bits^n$, and define propositional formula
$$
\tau_b(g_n)
$$
expressing $\forall x \in \bits^n g_n(x) \neq b$. The formula uses $n$ atoms for bits of $x$
and further $poly(n)$ atoms for bits of the computation $y$ of $g_n$ on $x$ and says, in a
DNF form, that either $y$ is not a valid computation on input $x$ or the output of the computation differs from 
$b$. Clearly
$$
\tau_b(g_n) \in \mbox{TAUT}\ \Leftrightarrow\ 
b \notin Rng(g_n)\ .
$$
These formulas were defined in \cite{ABRW,Kra-wphp} and lead to the theory of proof complexity generators 
proposing several candidate tautologies of the form above
as possibly hard for strong (or all) proof systems. The reader may find an
overview of the theory in \cite[Chpts.29 and 30]{k2} (no need to read the first 28 chapters).

\bigskip
\noindent
{\bf Acknowledgements:}

I thank Sam Buss, Bruce Kapron, Igor C.~Oliveira and Jan Pich for comments on earlier versions of this paper.

\end{document}